\documentclass[
preprint,
 amsmath,amssymb,
 aps,
prc,
]{revtex4-1}
\usepackage{graphicx}
\usepackage{braket}
\usepackage{longtable}

\begin{document}


\title{Gamow-Teller transitions and proton-neutron pair correlation in $N=Z=$ odd $p$-shell nuclei}


\author{Hiroyuki Morita and Yoshiko Kanada-En'yo}
\affiliation{Department of Physics, Kyoto University, Kyoto 606-8502, Japan.}

\date{\today}

\begin{abstract}
We have studied the Gamow-Teller (GT) transitions from $N=Z+2$ neighbors to $N=Z=$ odd nuclei in $p$-shell region
by using isospin-projected and $\beta\gamma$-constraint antisymmetrized molecular dynamics combined with generator coordinate method.
The calculated GT transition strengths from $0^+1$ states to $1^+0$ states 
such as ${}^{6} \textrm{He}(0_1^+1)\rightarrow{}^{6} \textrm{Li}(1_1^+0)$, ${}^{10} \textrm{Be}(0_1^+1)\rightarrow{}^{10} \textrm{B}(1_1^+0)$, 
and ${}^{14} \textrm{C}(0_1^+1)\rightarrow{}^{14} \textrm{N}(1_2^+0)$ exhaust more than 50\% of the sum rule.
These $N=Z+2$ initial states and $N=Z=$ odd final states are found to dominantly have $S=0,T=1$ $nn$ pairs and $S=1,T=0$ $pn$ pairs, respectively.
Based on two-nucleon ($NN$) pair picture, we can understand the concentration of the GT strengths as
the spin-isospin-flip transition $nn(S=0,T=1)\rightarrow pn(S=1,T=0)$ in $LS$-coupling scheme.
The GT transition can be a good probe to identify the spin-isospin partner states with $nn$ pairs and $pn$ pairs of $N=Z+2$ and $N=Z=$ odd nuclei, respectively.
\end{abstract}


\maketitle



\section{Introduction}
Proton and neutron ($pn$) correlation is one of the key phenomena to understand properties
of $N=Z=$ odd nuclei (see Ref.~\cite{PN_rev} and references therein).
In particular, a deuteron-like $T=0$ $pn$ pair plays an important role in low-lying states of 
light $N=Z=$ odd nuclei. 
Recently, a three-body model calculation of two nucleons with a doubly magic core nucleus
has been performed to study low-lying states of $N=Z=$ odd nuclei,
and the result indicates that deuteron-like $S=1,T=0$ and di-neutron-type 
$S=0,T=1$ pairs ($LS$-coupling $pn$ pairs) are predominantly formed at the surface of double magic cores such as ${}^{16} \textrm{O}$ \cite{Sagawa_three_body}. 
Moreover, in our previous work, we have studied 
$pn$ correlation in ${}^{10} \textrm{B}$ and found the low-lying $T=0$ and $T=1$ states dominantly have 
the $S=1,T=0$ and $S=0,T=1$ pairs around a $2\alpha$ core, respectively \cite{tbgAMD}.
It indicates that $LS$-coupling scheme is better than $jj$-coupling scheme 
to understand $pn$ pairs in light $N=Z=$ odd nuclei even though 
the $LS$-coupling $pn$ pairs may change to $jj$-coupling $pn$ pairs 
in heavy-mass systems because of the spin-orbit mean potential. 

Gamow-Teller transition is one of the useful observables to verify the $LS$-coupling $pn$ pairs 
because it is sensitive to the spin-isospin configuration.
For light $N=Z=$ odd nuclei, the GT operator flips nucleon spins and isospins of a pair 
and changes the di-neutron-type $nn$ pair to the deuteron-like $pn$ pair.
This type of GT transitions correspond to the collectivity of the proton-neutron pair and 
dominate the GT sum rule if the core parts are spin-isospin saturated systems and give 
no contribution to the GT transition. These modes are different from 
so-called Gamow-Teller giant resonances which are contributed by collectivity of the excess neutrons.
Recently, the super-allowed GT transitions in the low-energy region have been observed
and discussed in relation to the $pn$ correlations \cite{Fujita_Sc}.
The $LS$-coupling $pn$ pairs may play an important role to the low-energy super-allowed GT transitions.
However, there are a few theoretical works to systematically investigate 
$LS$-coupling $pn$ pairs in light $N=Z=$ odd nuclei,
though proton-neutron pairing correlations in medium and heavy mass $N=Z=$ odd nuclei 
have been discussed in $jj$-coupling scheme with mean field approaches \cite{Gez_Mixed_Pairing,Cederwall_pn,Qi_pn,HFB_pn_1,Yoshida_pn_1}.

The authors have a constructed new framework, isospin-projected $\beta\gamma$-constraint antisymmetrized
molecular dynamics ($T\beta\gamma$-AMD), which is useful in description of a $pn$ pair in deformed or clustered systems \cite{tbgAMD}. 
In this paper, we investigate the GT transitions and $pn$ pairs in $p$-shell nuclei
applying $T\beta\gamma$-AMD to ${}^{6} \textrm{Li}$, ${}^{10} \textrm{B}$, and ${}^{14} \textrm{N}$.
We discuss strong GT transitions in terms of $NN$ pair in $LS$-coupling scheme and 
propose an interpretation of the initial and final states as spin-isospin partners. 
A particular attention is paid on the role of non-zero intrinsic spin ($S=1$) of the $T=0$ $pn$ pair
and its coupling with the orbital angular momentum of $pn$ center of mass motion and that of core rotation. 

The paper is organized as follows. 
We briefly explain our framework in Sect.~\ref{method}.
We show the results of nuclear properties of energy spectra, $B(M1)$, $B(E2)$ and $B(\textrm{GT})$ in Sect.~\ref{results}.
We discuss the strong GT transitions in terms of $NN$ pair in $LS$-coupling scheme in Sect.~\ref{discussion} by analyzing the obtained wavefunctions.
A summary and an outlook are given in Sect.~\ref{summary}.

\section{Method}
\label{method}
\subsection{$T\beta\gamma$-AMD}
For $N=Z=$ odd nuclei, we apply $T\beta\gamma$-AMD \cite{tbgAMD}
in order to deal with the $pn$ pair formation as well as nuclear deformation and clustering. 
For $N=Z+2$ nuclei, we use $\beta\gamma$-constraint AMD \cite{bgAMD}, which has been used
for structure studies of light neutron-rich nuclei as well as $Z=N=$ even nuclei. 
We here briefly explain the formulation of $T\beta\gamma$-AMD.
Details of two methods, $T\beta\gamma$-AMD and $\beta\gamma$-AMD are described in Refs.~\cite{tbgAMD,bgAMD}.

In the original framework of AMD, a basis wavefunction is written by a Slater determinant of Gaussian wave packets,
\begin{equation}
\Ket{\Phi\left(\beta,\gamma\right)} = \mathcal{A} \left[\Ket{\phi_1} \Ket{\phi_2} \cdots \Ket{\phi_A}\right],
\end{equation}
\begin{equation}
\Ket{\phi_i} = \left(\frac{2\nu}{\pi}\right)^{\frac{3}{4}} \exp\left[-\nu\left(\bm{r}_i-\frac{\bm{Z}_i}{\sqrt{\nu}}\right)^2\right] \Ket{\bm{\xi}_i}\Ket{\tau_i}.
\end{equation}
In the present work, we use $\nu = 0.235$ for $p$-shell nuclei as used in Refs.~\cite{pshell_0235_1,pshell_0235_2,bgAMD,pshell_0235_4,pshell_0235_5,pshell_0235_6}.
In $T\beta\gamma$-AMD, we perform parity and isospin (${}^{\pi}T$) projections before variation as
\begin{equation}
\Ket{\Phi^{\pi T}\left(\beta,\gamma\right)} = \hat{P}^\pi\hat{P}^T \Ket{\Phi\left(\beta,\gamma\right)},
\end{equation}
where $\hat{P}^\pi$ and $\hat{P}^T$ are parity projection operator and isospin projection operator, respectively.
For the ${}^{\pi}T$-projected AMD wavefunction, 
we perform energy variation under the constraint on quadrupole deformation parameters 
$\beta\gamma$
and obtain the optimum solution for each set of $\beta$ and $\gamma$ values. 
In order to obtain wavefunctions for the $n$th $J^\pi T$ state (denoted by $J^\pi_n T$), 
we superpose the angular momentum eigenstates projected from 
the obtained wavefunctions $\Ket{\Phi^{\pi T}\left(\beta_i,\gamma_i\right)}$, 
\begin{equation}
\Ket{J^\pi_n T;M} = \sum_{iK} c^{iK}_n \hat{P}_{MK}^J \Ket{\Phi^{\pi T}\left(\beta_i,\gamma_i\right)},
\end{equation}
where $\hat{P}_{MK}^J$ is the angular momentum projection operator.
Here, the parameters, $\beta$ and $\gamma$, are treated as generator coordinates in the generator coordinate method (GCM), and the $K$-mixing is taken into account. 
We call this method $T\beta\gamma$-AMD+GCM. 

\subsection{Effective interactions}
We use the Hamiltonian
\begin{equation}
H = K-K_\textrm{cm} + V_\textrm{c} + V_{LS} + V_\textrm{Coulomb},
\end{equation}
where $K$ is the kinetic energy, $K_\textrm{cm}$ is the kinetic energy of the center of mass,
and $V_\textrm{c}$, $V_{LS}$, and $V_\textrm{Coulomb}$ are the central, spin-orbit, and Coulomb forces, respectively. 
For the central and spin-orbit forces, we use the effective nucleon-nucleon ($NN$) forces
same as those used for ${}^{10} \textrm{B}$ in the previous work \cite{tbgAMD}.
Namely, we use the Volkov No.~2 force of the central force with the Majorana exchange parameter $m=0.6$ 
and the G3RS force of the spin-orbit force with the strength parameters $u_1=-u_2=1300$ MeV.

For the Bartlett and Heisenberg parameters, $b$ and $h$, of the Volkov No.~2 force,
we use $b=h=0.125$ for ${}^{6} \textrm{Li}$, which reproduce the $S$-wave $NN$ scattering lengths in the $T=0$ and $T=1$ channels.
For ${}^{10} \textrm{B}$ and ${}^{14} \textrm{N}$, we adopt a parameterization
$b=h=0.06$ phenomenologically modified 
so as to describe energy difference between the lowest $T=0$ and $T=1$ states in each nucleus. 
The parameters $b$ and $h$ 
control the ratio ($f$) of the central force in the $T=0$ channel to that in the $T=1$ channel. 
The present choices, $b=h=0.125$ and $b=h=0.06$, give the ratios $f=1.67$ and $1.27$, respectively. 
The decrease of $f$ is consistent with the naive expectation that the $T=0$ interaction is somewhat suppressed by a nuclear medium effect.
We should comment that, even though relative position between $T=1$ and $T=0$ spectra
is sensitive to $b$ and $h$, we obtain almost the same energy spectra in each isospin channel
and also qualitatively similar results for structure properties of ${}^{10} \textrm{B}$ and ${}^{14} \textrm{N}$ in the cases of $b=h=0.125$ and $b=h=0.06$. 

\section{Results}
\label{results}
\begin{figure*}
\includegraphics[width=0.75\hsize]{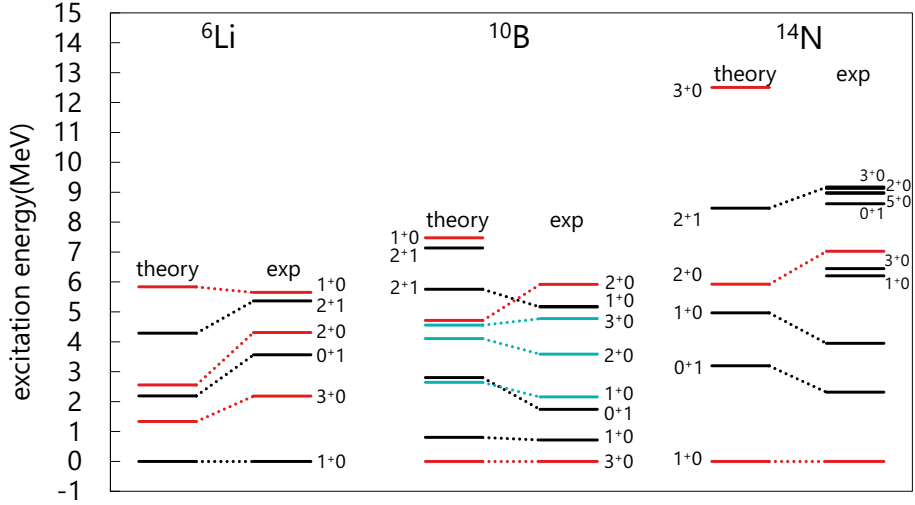}
\caption{Spectra of ${}^{6} \textrm{Li}$, ${}^{10} \textrm{B}$, and ${}^{14} \textrm{N}$ calculated by 
$T\beta\gamma$-AMD+GCM and those of the experimental data \cite{ExpSpec6,ExpSpec,ExpSpec4}.}
\label{figure1}
\end{figure*}

Calculated energy spectra of ${}^{6} \textrm{Li}$, ${}^{10} \textrm{B}$, and ${}^{14} \textrm{N}$ obtained by $T\beta\gamma$-AMD+GCM are shown in Fig.~\ref{figure1} compared with experimental spectra.
The present calculation reasonably reproduces the low-energy spectra of these nuclei.

The calculated binding energies, magnetic dipole moments ($\mu$), electric quadrupole moments ($Q$), 
and $E2$ and $M1$ transition strengths of ${}^{6} \textrm{Li}$, ${}^{10} \textrm{B}$, and ${}^{14} \textrm{N}$ are listed in Table~\ref{property} together with experimental data.
The present calculation quantitatively or qualitatively reproduces the experimental data of these properties.

$\mu$ moments and $B(M1)$ as well as the 
 GT transition strengths are observables that sensitively reflect spin configurations.
The calculated $\mu$ moments of the ground states, ${}^{6} \textrm{Li}(1^+_10)$, ${}^{10} \textrm{B}(3^+_10)$, and ${}^{14} \textrm{N}(1^+_10)$, and that of ${}^{10} \textrm{B}(1^+_10)$ agree well to the experimental data.
For the $M1$ transition in ${}^{6} \textrm{Li}$, the remarkably large $B(M1;0^+_11\to 1^+_10)$ is well reproduced by the calculation. 
For ${}^{10} \textrm{B}$, the $M1$ transitions between $T=1$ and $T=0$ states are qualitatively described 
although quantitative reproduction is not satisfactory in the present calculation.
For ${}^{14} \textrm{N}$, the present calculation describes the general trend of the relatively strong $M1$ transitions for $1^+_20\to 0^+_11$ and $2^+_11\to 2^+_10$ compared with those for other transitions.

For ${}^{10} \textrm{B}$, the calculated $Q$ moment and $B(E2;3^+_20\to 1^+_10)$ are large consistently with the experimental data because of the prolate deformation. 
The prolate deformation of ${}^{10} \textrm{B}$ is caused by formation of a $2\alpha$ core as shown later. 

\begin{longtable}{ccc}
\caption{
Binding energies, $\mu$ and $Q$ moments, 
and $M1$ and $E2$ transition strengths of ${}^{6} \textrm{Li}$, ${}^{10} \textrm{B}$, and ${}^{14} \textrm{N}$.
The calculated values obtained by $T\beta\gamma$-AMD+GCM are shown.
Experimental data are taken from \cite{ExpSpec6,ExpSpec,ExpSpec4}.}
\label{property}
\\
\hline\hline
Observable&$T\beta\gamma$-AMD+GCM&Exp\\
\hline
\endfirsthead
\hline\hline
Observable&$T\beta\gamma$-AMD+GCM&Exp\\
\hline
\endhead
\hline\hline
\endfoot
\hline\hline
\endlastfoot
${}^{6} \textrm{Li}$\\
\hline
$\left|E(1_1^+0)\right|$ (MeV) & 29.55 & 31.99\\
$\mu(1_1^+0)$ ($\mu_N$) &0.87 &0.82 \\
$Q(1_1^+0)$ ($e$~$\textrm{fm}^2$) &0.09 &-0.08 \\
$B(E2;3_1^+0\rightarrow1_1^+0)$&	3.79&	10.69	\\
$B(M1;0_1^+1\rightarrow1_1^+0)$&	13.73&	15.43	\\
$B(E2;2_1^+0\rightarrow1_1^+0)$&	5.15&	4.40	\\
$B(M1;2_1^+1\rightarrow1_1^+0)$&	0.01&	0.15	\\
\hline
${}^{10} \textrm{B}$\\
\hline
$\left|E(3_1^+0)\right|$ (MeV) & 60.35 & 64.75\\
$\mu(3_1^+0)$ ($\mu_N$) & 1.83 &1.80\\
$\mu(1_1^+0)$ ($\mu_N$) & 0.84 &0.63\\
$Q(3_1^+0)$ ($e$~$\textrm{fm}^2$) & 8.45 &8.47\\
$B(E2;1_1^+0\rightarrow3_1^+0)$&	4.03&	4.15	\\
$B(M1;0_1^+1\rightarrow1_1^+0)$&	14.98&	7.52	\\
$B(M1;1_2^+0\rightarrow0_1^+1)$&	0.05&	0.19	\\
$B(E2;1_2^+0\rightarrow1_1^+0)$&	9.23&	15.61	\\
$B(E2;1_2^+0\rightarrow3_1^+0)$&	2.02&	1.70	\\
$B(E2;2_1^+0\rightarrow3_1^+0)$&	0.34&	1.15	\\
$B(E2;3_2^+0\rightarrow1_1^+0)$&	10.56&	19.71	\\
$B(M1;2_1^+1\rightarrow2_1^+0)$&	1.84&	2.52	\\
$B(M1;2_1^+1\rightarrow1_2^+0)$&	2.60&	3.06	\\
$B(M1;2_1^+1\rightarrow1_1^+0)$&	0.31&	0.32	\\
\pagebreak\hline
${}^{14} \textrm{N}$\\
\hline
$\left|E(1_1^+0)\right|$ (MeV) & 108.60 & 104.66\\
$\mu(1_1^+0)$ ($\mu_N$) & 0.34 &0.40 \\
$Q(1_1^+0)$ ($e$~$\textrm{fm}^2$) &0.53 &1.93 \\
$B(M1;0_1^+1\rightarrow1_1^+0)$&	0.76&	0.05	\\
$B(M1;1_2^+0\rightarrow0_1^+1)$&	3.72&	1.79	\\
$B(E2;1_2^+0\rightarrow1_1^+0)$&	3.25&	4.41	\\
$B(E2;2_1^+0\rightarrow1_1^+0)$&	2.95&	3.61	\\
$B(M1;2_1^+1\rightarrow2_1^+0)$&	4.65&	1.74	\\
$B(M1;2_1^+1\rightarrow1_1^+0)$&	0.00&	0.59	\\
\end{longtable}

In order to calculate GT transition strengths,
we apply $\beta\gamma$-AMD and obtain wavefunctions for the ground and excited states of ${}^{6} \textrm{He}$, 
${}^{10} \textrm{Be}$, and ${}^{14} \textrm{C}$, which are isobaric analogue states of 
$T=1$ states of ${}^{6} \textrm{Li}$, ${}^{10} \textrm{B}$, and ${}^{14} \textrm{N}$.
Table~\ref{GTstrength} shows the calculated $B(\textrm{GT})$:
\begin{equation}
\label{GTdef}
B\left(\textrm{GT}\right) = \frac{1}{2J_i+1}\left|\Braket{J_f||\sum_i\bm{\sigma}^i\tau^i||J_i}\right|^2.
\end{equation}
In the present paper, we define $B(\textrm{GT})$ by matrix elements of the spin and isospin operators without the factor $(g_A/g_V)^2$.
For all low-lying states of ${}^{6}\textrm{Li}$, ${}^{10}\textrm{B}$, and ${}^{14}\textrm{N}$, we find $T=0$ states that have strong GT transitions with large percentages of the sum rule $\sum B(\textrm{GT})=3(N-Z)=6$. 
These final states in $Z=N=$ odd nuclei can be regarded as ``spin-isospin partners'' of the corresponding $T=1$ initial states 
because they are approximately spin-isospin-flipped states having spatial configurations similar to the initial states. 
The concept of the spin-isospin partners is an extension of isobaric analogue state (IAS) to the GT transition.
The assignments of the spin-isospin partners in the following discussions are based on the calculated GT transition strengths 
and also spin and orbital configurations in $LS$-coupling scheme of $NN$ pairs.

For ${}^{6} \textrm{Li}$, the GT transition from the ground state ${}^{6}\textrm{He} (0^+_11)$ to $1^+_10$ exhausts a large fraction of the sum rule consistently with the experimental data, 
whereas that to $1^+_20$ is weak. 
This fact indicates that the ground states of ${}^{6} \textrm{Li}$ and ${}^{6}\textrm{He}$ are almost ideal spin-isospin partners. 
For the GT transitions from the excited state, ${}^{6}\textrm{He} (2^+_11)$, 
we obtain strong transitions to $1^+_20$, $2^+_10$, and $3^+_10$.
The summation of $B(\textrm{GT})$ values for these three states is about 50\% of the sum rule, 
and therefore, these states are regarded as the set of spin-isospin partners with $J^\pi=\{1^+,2^+,3^+\}$ of the ${}^{6}\textrm{He} (2^+_11)$.

Also for ${}^{10} \textrm{B}$, the GT transition strength from the ground state ${}^{10} \textrm{Be}(0^+_11)$ is concentrated to $1^+_10$.
For the GT transitions from the excited states ${}^{10} \textrm{Be}(2^+_11)$ and ${}^{10} \textrm{Be}(2^+_21)$, significant strengths 
are obtained for transitions to $1^+_20$, $1^+_30$, $2^+_10$, $2^+_20$, $3^+_10$, and $3^+_20$, which can be assigned to 
two sets of spin-isospin partners with $J^\pi=\{1^+,2^+,3^+\}$ of ${}^{10} \textrm{Be}(2^+_11)$ and ${}^{10} \textrm{Be}(2^+_21)$. 
In particular, the transitions from ${}^{10} \textrm{Be}(2^+_11)$ are significantly strong 
to $1^+_20$ and $3^+_20$ indicating that these states are regarded as spin-isospin partner states. 
In the transitions from ${}^{10} \textrm{Be}(2^+_21)$, the  strengths to $1^+_30$ and $3^+_10$ are significantly large, and hence, 
these states can be assigned to the spin-isospin partners of ${}^{10} \textrm{Be}(2^+_21)$.
For assignment of $J^\pi=2^+$ states, the strengths $2^+_11\to 2^+_10$, $2^+_11\to 2^+_20$, $2^+_21\to 2^+_10$, and $2^+_21\to 2^+_20$ are comparable. 
It indicates that two $J^\pi=2^+$ states corresponding to the partner states of ${}^{10} \textrm{Be}(2^+_11)$ and ${}^{10} \textrm{Be}(2^+_21)$ are strongly mixed with each other.

For ${}^{14} \textrm{N}$, the strongest GT transition from the ground state ${}^{14} \textrm{C}(0^+_11)$ is obtained for $1^+_20$ consistently with the experimental data.
It indicates that not $1_1^+0$ but $1_2^+0$ of ${}^{14}\textrm{N}$
is the spin-isospin partner in the $A=14$ systems.
Compared with the dominant transition $0^+_11\rightarrow1^+_20$, 
the calculated transition $0^+_11\rightarrow1^+_10$ is relatively minor 
but it does not reproduce the anomalously small value of the experimental datum.
For the transitions from the first excited state $2^+_11$, strengths to $1^+_10$, $2^+_10$, and $3^+_10$ are significant and the sum of them exhausts more than 80\% of the sum rule; thus, these states are regarded as the spin-isospin partners. \newpage

\begin{longtable}{ccc}
\caption{
Calculated values of $B(\textrm{GT})$ defined in Eq.~(\ref{GTdef}) are shown.
Experimental data are taken from \cite{ExpSpec3,ExpSpec4,ExpSpec5,GT_14C_1,GT_14C_2}.
The value in the parenthesis is 
the experimental datum for the mirror transition ${}^{10} \textrm{C}(0_1^+1)\rightarrow{}^{10} \textrm{B}(1_1^+0)$.
}
\label{GTstrength}
\\
\hline\hline
Initial$\rightarrow$Final&$T\beta\gamma$-AMD+GCM&Exp\\
\hline
\endfirsthead
\hline\hline
Initial$\rightarrow$Final&$T\beta\gamma$-AMD+GCM&Exp\\
\hline
\endhead
\hline\hline
\endfoot
\hline\hline
\endlastfoot
${}^{6} \textrm{He}\rightarrow{}^{6} \textrm{Li}$\\
\hline
$0_1^+1\rightarrow1_1^+0$&	5.31&	3.02	\\ 
$0_1^+1\rightarrow1_2^+0$&	0.00&	--	\\
$2_1^+1\rightarrow1_1^+0$&	0.01&	--	\\
$2_1^+1\rightarrow3_1^+0$&	0.97&	--	\\
$2_1^+1\rightarrow2_1^+0$&	1.00&	--	\\
$2_1^+1\rightarrow1_2^+0$&	1.10&	--	\\
\hline
${}^{10} \textrm{Be}\rightarrow{}^{10} \textrm{B}$\\
\hline
$0_1^+1\rightarrow1_1^+0$&	4.95&	(2.20)	\\ 
$0_1^+1\rightarrow1_2^+0$&	0.15&	--	\\
$0_1^+1\rightarrow1_3^+0$&	0.00&	--	\\
$2_1^+1\rightarrow3_1^+0$&	0.63&	0.07	\\ 
$2_1^+1\rightarrow1_1^+0$&	0.06&	--	\\
$2_1^+1\rightarrow1_2^+0$&	0.81&	--	\\
$2_1^+1\rightarrow2_1^+0$&	0.77&	--	\\
$2_1^+1\rightarrow3_2^+0$&	1.71&	--	\\
$2_1^+1\rightarrow1_3^+0$&	0.26&	--	\\
$2_1^+1\rightarrow2_2^+0$&	0.86&	--	\\
$2_2^+1\rightarrow3_1^+0$&	1.54&	0.85	\\ 
$2_2^+1\rightarrow1_1^+0$&	0.01&	--	\\
$2_2^+1\rightarrow1_2^+0$&	0.23&	--	\\
$2_2^+1\rightarrow2_1^+0$&	0.71&	--	\\
$2_2^+1\rightarrow3_2^+0$&	0.26&	--	\\
$2_2^+1\rightarrow1_3^+0$&	0.82&	--	\\
$2_2^+1\rightarrow2_2^+0$&	0.79&	--	\\
\pagebreak\hline
${}^{14} \textrm{C}\rightarrow{}^{14} \textrm{N}$\\
\hline
$0_1^+1\rightarrow1_1^+0$&	0.30&	$3.64\times10^{-6}$	\\ 
$0_1^+1\rightarrow1_2^+0$&	4.32&	1.70	\\ 
$2_1^+1\rightarrow1_1^+0$&	1.13&	0.17	\\ 
$2_1^+1\rightarrow2_1^+0$&	1.77&	--	\\
$2_1^+1\rightarrow3_1^+0$&	2.35&	--	\\
\end{longtable}

\section{Discussion}
\label{discussion}
In the previous section, we have discussed the assignments of spin-isospin partners 
focusing on the strong GT transitions in $A=6$, $A=10$, and $A=14$ nuclei.
In this section, we discuss detailed features of $NN$ pairs in the spin-isospin-partner states.

\subsection{Intrinsic structure and spatial distribution of a proton-neutron pair}
\begin{figure}
\includegraphics[width=0.5\hsize]{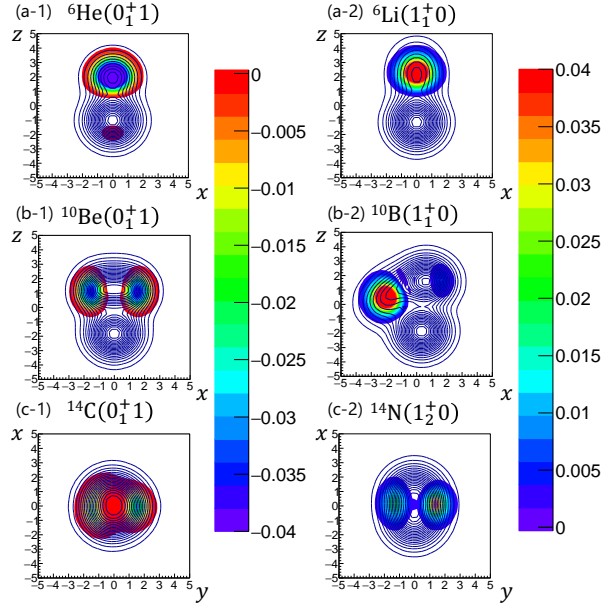}
\caption{
The colored contours show two-nucleon-pair density $\rho_{NN}(\bm{r})$ of (a-1)${}^{6} \textrm{He}(0_1^+1)$, (a-2)${}^{6} \textrm{Li}(1_1^+0)$, (b-1)${}^{10} \textrm{Be}(0_1^+1)$, (b-2)${}^{10} \textrm{B}(1_1^+0)$, (c-1)${}^{14} \textrm{C}(0_1^+1)$, and (c-2)${}^{14} \textrm{N}(1_2^+0)$.
The blue contours show the one-body density distribution $\rho(\bm{r})$.
}
\label{figure3}
\end{figure}

In the obtained wavefunctions for the $A=6$, $A=10$, and $A=14$ systems, 
$NN$ pairs are found to be formed around $\alpha$, $2\alpha$, and ${}^{12} \textrm{C}$ cores, respectively. 
In order to see the spatial distribution of the $S=1,T=0$ and $S=0,T=1$ $NN$ pairs in the spin-isospin partners, 
we calculate two-particle density ${\rho}_{ST}(\bm{r})$ at the identical point $\bm{r}$ in the intrinsic states defined as 
\begin{eqnarray}
\rho_{ST}(\bm{r}) = \frac{\Braket{\Phi^T(\beta,\gamma)|\hat{\rho}_{ST}(\bm{r})|\Phi^T(\beta,\gamma)}}{\Braket{\Phi^T(\beta,\gamma)|\Phi^T(\beta,\gamma)}},\\
\hat{\rho}_{ST}(\bm{r}) \equiv \sum_{ij}\hat{P}^S_{ij}\hat{P}^T_{ij}\delta(\bm{r}-\hat{\bm{r}}_i)\delta(\bm{r}-\hat{\bm{r}}_j),
\end{eqnarray}
where $\hat{P}^S_{ij}$ and $\hat{P}^T_{ij}$ are the spin and isospin projection operators for two particles. 
We define the two-nucleon-pair density 
$\rho_{NN}(\bm{r})\equiv\rho_{10}(\bm{r})-\rho_{01}(\bm{r})$ to cancel $NN$ pair contributions from $\alpha$ clusters which contain the 
same numbers of $S=1,T=0$ $NN$ pairs as those of $S=0,T=1$ $NN$ pairs.
With this definition, positive (negative) regions of $\rho_{NN}(\bm{r})$ indicate $S=1,T=0$ ($S=0,T=1$) $pn$-pair distributions in $T=0$ ($T=1$) states.
In Fig.~\ref{figure3}, we show the two-nucleon-pair density $\rho_{NN}(\bm{r})$ together 
with the one-body density distribution in the single Slater-determinant state 
which has the largest overlap in the $\beta\gamma$ plane with the wavefunction
for each of the ground $0^+_11$ states of ${}^{6}\textrm{He}$, ${}^{10}\textrm{Be}$, and ${}^{14}\textrm{C}$ and their spin-isospin partner $1^+0$ states of the $N=Z=$ odd nuclei. 

In the ground state of ${}^{6} \textrm{Li}$, an $\alpha$ particle and a $T=0$ $pn$ pair are formed as seen in Fig.~\ref{figure3}(a-2).
The $pn$ pair spatially develops away from the $\alpha$ core and it shows deuteron-like nature.
Also in ${}^{6} \textrm{He}$, the two-neutron pair appears around an $\alpha$ core (Fig.~\ref{figure3}(a-1)).
The spatial distribution of the $nn$ pair density in ${}^{6} \textrm{He}$ is quite similar to that of 
the $pn$ pair density in ${}^{6} \textrm{Li}$
indicating that these states are good spin-isospin partner states, 
in which the two-nucleon spin $S$ and isospin $T$ flip from $nn(T=1, S=0)$ in ${}^{6} \textrm{He}$ to $pn(T=0, S=1)$ in ${}^{6} \textrm{Li}$.

In ${}^{10} \textrm{B}(1_1^+0)$, two $\alpha$ clusters and a $T=0$ $pn$ pair are formed (see Fig.~\ref{figure3}(b-2)). 
The $T=0$ $pn$ pair develops away from the $2\alpha$ core similarly to ${}^{6} \textrm{Li}$, 
whereas the $nn$ pair in ${}^{10} \textrm{Be}(0_1^+1)$ is not so developed spatially 
but is distributed at the nuclear surface showing a feature of two $p$-orbit neutrons (Fig.~\ref{figure3}(b-1)).
Although the single Slater-determinant state with the largest overlap for ${}^{10} \textrm{Be}(0_1^+1)$ shows less development 
of the two-nucleon pair than that for ${}^{10} \textrm{B}(1^+_10)$, 
however, in the $\beta\gamma$-AMD+GCM result, the spatially developed $nn$ pair components are largely mixed 
because the two-neutron pair can move away from the $2\alpha$ core along a plateau 
toward a finite $\gamma$ region in the $J^\pi=0^+$ energy surface of ${}^{10} \textrm{Be}$. 
As a result, the $nn$ pair distribution in ${}^{10} \textrm{Be}(0_1^+1)$ 
has large overlap with the $pn$ pair distribution in ${}^{10} \textrm{B}$,
and therefore, these states have the strong GT transition and are regarded as the partner states.

In ${}^{14} \textrm{N}(1^+_20)$ (see Fig.~\ref{figure3}(c-2)), 
a $T=0$ $pn$ pair is distributed at the surface of the oblately deformed ${}^{12} \textrm{C}$ core.
In ${}^{14} \textrm{C}(0^+_11)$, the $nn$ pair density around the ${}^{12} \textrm{C}$ core 
shows distribution similar to the $pn$ pair in ${}^{14} \textrm{N}(1^+_20)$.
The $NN$ pairs in ${}^{14} \textrm{N}(1^+_20)$ and ${}^{14} \textrm{C}(0^+_11)$
show no spatial development and dominantly consist of $p$-orbit nucleons. 
If we consider a ${}^{16} \textrm{O}$ core, these states can be understood as two-hole pairs in the $p$-shell of the ${}^{16} \textrm{O}$ core. 

Let us discuss spatial development of the $NN$ pairs with $A$ increasing in the $A=6$, $A=10$, and $A=14$ systems.
In the $0^+1$ ground states of $N=Z+2$ nuclei, the $nn$ pair is mostly developed spatially in the $A=6$ nucleus and 
comes down to the $p$-shell configurations in $A=10$ and $A=14$ nuclei with increase of the mass number.
In the partner $1^+$ states of the $N=Z=$ odd nuclei, the spatially developed $T=0$ $pn$ pair is prominent in the $A=6$ nucleus and it 
more or less weakens but still remains even in the $A=10$ nucleus, 
and finally comes down to the $p$-shell configuration in the $A=14$ nucleus.
This result reflects the feature that the $T=0$ $pn$ pairs in $N=Z=$ odd nuclei are robuster than $nn$ pairs in $N=Z+2$ nuclei.
Indeed, the $T=0$ $pn$ pairs are described well by $LS$-coupling scheme, whereas 
$nn$ pairs are somewhat broken from $LS$-coupling scheme and contain 
mixing of $jj$-coupling components, in particular, in the $A=10$ and $A=14$ nuclei as shown later in analysis of spin configurations.

\subsection{$pn$ pairs in $LS$-coupling scheme and spin-isospin partners}
To quantitatively discuss the spin and orbital configurations, 
we show the expectation values of the squared intrinsic spin and orbital angular momentum ($\Braket{\bm{S}^2}$ and $\Braket{\bm{L}^2}$) in Table~\ref{LLSS}.
Note that $\Braket{\bm{S}^2}$ approximately indicates the expectation value of 
the squared intrinsic spin of a $NN$ pair around a core because core contribution is minor in the present case:
the obtained states of the $A=6$ and $A=10$ nuclei are understood by two particles around $S=0$ cores such as $\alpha$ and $2\alpha$ 
and those of the $A=14$ nuclei are approximately interpreted as two-hole states of ${}^{16} \textrm{O}$.
$T=1$ states of the $N=Z=$ odd nuclei have almost same expectation values as those of the $N=Z+2$ nuclei 
because they are isobaric analogue states.

In the obtained states of the $A=6$, $A=10$, and $A=14$ nuclei, 
the spin expectation values of $T=0$ ($T=1$) states are close to the value 
$\Braket{\bm{S}^2}=2$ ($\Braket{\bm{S}^2}=0$) for $S=1$ ($S=0$) component.
It implies that $LS$-coupling $NN$ pairs are formed as leading components in particular in light nuclei.
As the mass number increases, the $T=1, S=0$ $NN$ pairs in $LS$-coupling scheme are somewhat broken 
into $jj$-coupling pairs because of the spin-orbit mean potential. 
We can see this systematics especially in $\Braket{\bm{S}^2}$ for the $0_1^+1$ states. 
${}^{6} \textrm{He}(0_1^+1)$ has almost pure $S=0$ component with only 6\% mixing of $S=1$ component estimated from $\Braket{\bm{S}^2}=0.12$.
However, ${}^{14} \textrm{C}(0_1^+1)$ has a broken $S=0$ two-hole pair with significant $S=1$ component up to 27\%. 
In contrast to the $T=1$ states, the $LS$-coupling $pn$ pairs in the $T=0$ states is not broken;
the $S=0$ mixing is found to be less than 6\% for all the $T=0$ states. 
This result implies that $T=0, S=1$ $pn$ pairs are robuster than $T=1, S=0$ $NN$ pairs. 
Even though $NN$ pairs are not necessarily ideal $LS$-coupling pairs, 
they have $LS$-coupling features as major components and can be qualitatively understood by 
$LS$-coupling scheme.

For the orbital angular momentum, values of $\Braket{\bm{L}^2}\approx 0$ and 
$\Braket{\bm{L}^2}\approx 6$ indicate dominant $L=0$ and $L=2$ components, respectively.
In $A=6$ nuclei, the total orbital angular momentum $L$ is contributed only by the orbital angular momentum $L_{NN}$ of the $NN$ pair 
because the $\alpha$ core is spherical.
Therefore, the ground states ${}^{6} \textrm{He}(0^+_11)$ and ${}^{6} \textrm{Li}(1^+_10)$ are understood well 
by $S=0,T=1$ and $S=1,T=0$ $NN$ pairs moving around the $\alpha$ in $L_{NN}=0$ wave, 
whereas the excited states ${}^{6} \textrm{He}(2^+_11)$ and ${}^{6} \textrm{Li}(1^+_20,2^+_10,3^+_10)$ contain $S=0,T=1$ and $S=1,T=0$ $NN$ pairs in $L_{NN}=2$ wave. 
In the $A=10$ nuclei, not only $L_{NN}$ but also collective rotation of the $2\alpha$ core with the orbital angular momentum $L_\textrm{core}$ contributes to $L$. 
For ${}^{10} \textrm{Be}(0^+_11)$ and 
${}^{10} \textrm{B}(1^+_10)$,
$\Braket{\bm{L}^2}\approx 0$ indicates that these states can be approximately described by the $S=0,T=1$ and $S=1,T=0$ $NN$ pairs with $L_{NN}=L_\textrm{core}=0$.
The orbital angular momentum $L\approx 2$ of ${}^{10} \textrm{Be}(2^+_11)$
mainly comes from the core rotation $L_\textrm{core}=2$, whereas that of 
${}^{10} \textrm{Be}(2^+_21)$ is contributed mainly by $L_{NN}=2$ from the $NN$ pair 
rotation because the former and the latter states are a member of the $K=0$ ground band
and that of 
the $K=2$ side band, respectively. 
It means that ${}^{10} \textrm{Be}(0^+_11)$ and ${}^{10} \textrm{Be}(2^+_21)$ are described 
by $S=0$ $nn$ pairs in $L_{NN}=0$ and $L_{NN}=2$ waves, respectively, 
and ${}^{10} \textrm{Be}(2^+_11)$ is understood by a $S=0$ $nn$ pair with the rotating $2\alpha$ core ($L_\textrm{core}=2$). 
The corresponding spin-isospin partners in ${}^{10} \textrm{B}$ should have $S=1,T=0$ $pn$ pairs with consistent spatial configurations. 
For the $A=14$ systems, the dominant components of ${}^{14} \textrm{C}(0^+_11)$ and ${}^{14} \textrm{N}(1^+_21)$
have two holes in ${}^{16} \textrm{O}$ coupled to be $S=0,T=1$ and $S=1,T=0$ pairs in $L_{NN}=0$ wave, whereas those of 
${}^{14} \textrm{C}(2^+_11)$ and ${}^{14} \textrm{N}(1^+_10,2^+_10,3^+_10)$ are understood by $S=0,T=1$ and $S=1,T=0$ two-hole pairs in $L_{NN}=2$ wave. 

\begin{figure*}
\includegraphics[width=0.85\hsize]{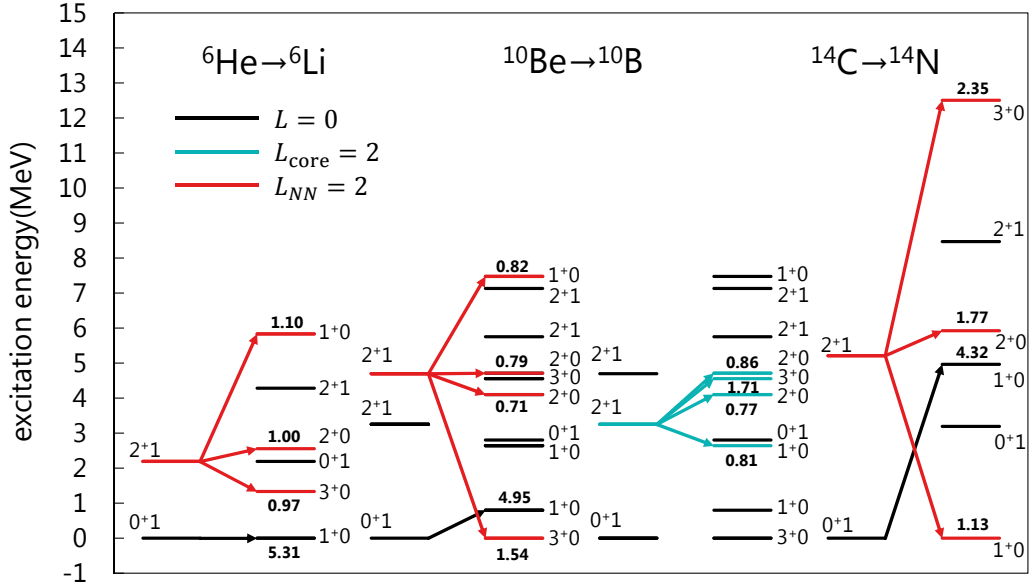}
\caption{
GT transitions ${}^{6} \textrm{He}\rightarrow{}^{6} \textrm{Li}$, ${}^{10} \textrm{Be}\rightarrow{}^{10} \textrm{B}$, and ${}^{14} \textrm{C}\rightarrow{}^{14} \textrm{N}$ calculated 
by $T\beta\gamma$-AMD+GCM.
}
\label{figure2}
\end{figure*}

Based on $LS$-coupling scheme of $NN$ pairs, 
we can easily understand spin-isospin partners and their strong GT transitions.
The GT operator changes intrinsic spin configuration with $\Delta S=1$
from $T=1$ states to $T=0$ states but it does not affect orbital configurations. 
In case of $L_\textrm{core}=0$, $nn$ pairs in $[L_{NN}=0, S_{NN}=0]_{J=0}$ initial states
change directly into $T=0$ $pn$ pairs in $[L_{NN}=0, S_{NN}=1]_{J=1}$ states with strengths of the sum rule value:
$\sum_n B(\textrm{GT}; 0^+1\rightarrow 1_n^+0)=6$
provided that core nuclei are spin-isospin saturated states and do not contribute to 
the GT transitions.
Similarly, we can easily understand the spin-isospin partners of $[L_{NN}=2, S_{NN}=0]_{J=2}$ initial states and $[L_{NN}=2, S_{NN}=1]_{J=1,2,3}$ final states.
Although $J$ in the final states is not unique because of angular momentum coupling of $S_{NN}=1$ with nonzero $L_{NN}$,
we can again obtain the sum rule: $\sum_{J=1,2,3} \sum_n B(\textrm{GT}; 2^+1\rightarrow J_n^+0)=6$.
It should be pointed out that, since $S=1$ $pn$ pairs in $L_{NN}=2$ wave feel spin-orbit mean 
potentials from core nuclei, energy spectra of the final $J=1,2,3$ states show spin-orbit splitting 
which plays an essential role to lower the $T=0$ states into the ground states in ${}^{10} \textrm{B}$ and ${}^{14} \textrm{N}$.
For deformed nuclei, we can also consider spin-isospin partners of
$[L_\textrm{core}=2, L_{NN}=0, S_{NN}=0]_{J=0}$ initial states and $[L_\textrm{core}=2, L_{NN}=0, S_{NN}=1]_{J=1,2,3}$ final states.

If the $NN$ pairs are broken into the $jj$-coupling pairs, the concentration of GT transition strengths
does not occur because initial states change into various $(jj')$ configurations.
In other words, the concentration of GT transition strengths to specific final states 
is a good measure for realization of $LS$-coupling $NN$ pairs. 
Based on the $LS$-coupling picture of $NN$ pairs, we assigned the spin-isospin partners 
for $T=1$ states and $T=0$ states 
with strong GT transition strengths which are qualitatively characterized by $\Delta T=0$, $\Delta S=1$, $\Delta L=0$ transitions.

Fig.~\ref{figure2} shows the calculated energy spectra and the GT transitions for 
the spin-isospin partners in $A=6$, $A=10$, and $A=14$. 
The GT transition from ${}^{6} \textrm{He}(0_1^+1)$ to ${}^{6} \textrm{Li}(1_1^+0)$ is enhanced 
because these states have the same $L_{NN}=0$ nature.
For the excited states, the GT transitions from ${}^{6} \textrm{He}(2_1^+1)$ to ${}^{6} \textrm{Li}(1_2^+0,2_1^+0,3_1^+0)$ are strong 
because of the transition from the $T=1, S=0$ pair to the $T=0, S=1$ pair in the dominant $L_{NN}=2$ component. 
The sum of the GT strengths from ${}^{6} \textrm{He}(2_1^+1)$ exhausts a large fraction of the sum rule value
indicating that the nature of spin-isospin partners still remains also in the excited states. 
In the energy spectra of ${}^{6} \textrm{Li}(1_2^+0,2_1^+0,3_1^+0)$, 
the ordering of $3_1^+0$, $2_1^+0$, and $1_2^+0$ is easily understood by the spin-orbit splitting for the $S=1$ $pn$ pairs in $L_{NN}=2$. 

We can also understand the GT transition from ${}^{10} \textrm{Be}(0_1^+1)$ to ${}^{10} \textrm{B}(1_1^+0)$ in the picture of $LS$-coupling $NN$ pairs 
as GT transition from a $nn$ pair to a $T=0$ $pn$ pair in $L_{NN}=0$.
For the excited states ${}^{10} \textrm{Be}(2_1^+1)$ and ${}^{10} \textrm{Be}(2_2^+1)$, 
two sets of $J^\pi=\{1^+,2^+,3^+\}$ for the spin-isospin partners appear 
in the $T=0$ spectra, 
but the $2^+0$ states are strongly mixed with each other because they almost degenerate energetically. 
${}^{10} \textrm{Be}(2_1^+1)$ has a rotating core with $L_\textrm{core}=2$ and it has strong 
transition strength to ${}^{10} \textrm{B}(1_2^+0,2_{1,2}^+0,3_2^+0)$, 
which almost degenerate because there is no spin-orbit splitting for the $T=0$ $pn$ pairs in $[L_\textrm{core}=2, L_{NN}=0, S_{NN}=1]_{J=1,2,3}$.
${}^{10} \textrm{Be}(2_2^+1)$ with a rotating $S=0$ $nn$ pair in $L_{NN}=2$ has dominant transition strength to ${}^{10} \textrm{B}(1_3^+0,2_{1,2}^+0,3_1^+0)$, 
which show large spin-orbit splitting of the $S=1$ $pn$ pairs in $L_{NN}=2$. 
As a result of the spin-orbit splitting, the $3^+0$ state partnered with ${}^{10} \textrm{Be}(2_2^+1)$ comes down to the ground state of ${}^{10} \textrm{B}$. 
This assignment is consistent with the experimental data of the strong GT transition for ${}^{10} \textrm{B}(3_1^+0)\to{}^{10}\textrm{Be}(2_2^+1)$ measured by charge exchange reactions \cite{ExpSpec5}.
Strictly speaking, it is in principle unable to definitely define $L_\textrm{core}$ and $L_{NN}$ for $N=Z=$ odd nuclei with deformed cores
because core nucleons and valence nucleons are identical fermions and indistinguishable in fully microscopic wavefunctions of identical fermions. 
Nevertheless, the GT transitions from $N=Z+2$ neighbors are observables and they enables us to classify the final states in $T=0$ $N=Z=$ odd nuclei in terms of $T=0$ $pn$ pairs in connection with $nn$ pairs in the initial states of $N=Z+2$ nuclei.

In ${}^{14} \textrm{N}$ spectra, low-lying states are understood as spin-isospin partners of 
${}^{14} \textrm{C}$ for $NN$ hole pairs in the ${}^{16} \textrm{O}$ core.
${}^{14} \textrm{C}(0_1^+1)$ has the strong GT transition not to the lowest $1^+0$ state but to the excited $1^+0$ state ${}^{14} \textrm{N}(1_2^+0)$ because these states have $NN$ hole pairs in the same $L_{NN}=0$ orbit. 
Then, the GT transition occurs from the $S=0$ $nn$ hole pair to the $S=1$ $pn$ hole pair. 
The GT transitions from ${}^{14} \textrm{C}(2_1^+1)$ to ${}^{14} \textrm{N}(1_1^+0,2_1^+0,3_1^+0)$ show spin-isospin-flip features of $NN$ hole pairs.
Indeed, ${}^{14} \textrm{N}(1_1^+0,2_1^+0,3_1^+0)$ spectra show the spin-orbit splitting 
of the $S=1$ $pn$ hole pairs in $L_{NN}=2$. 
Note that the ordering $1_1^+0$, $2_1^+0$, and $3_1^+0$ is opposite to that of the 
particle-particle pair case 
because the spin-orbit mean potentials for hole states are repulsive. 

Our assignments are consistent with the strong GT transition
for ${}^{14} \textrm{C}(0_1^+1)\to{}^{14} \textrm{N}(1_2^+0)$ 
experimentally measured by charge exchange reactions.
Moreover, for the transitions from ${}^{14} \textrm{N}(1_1^+0)$, 
relatively strong GT transitions to 
${}^{14} \textrm{C}(2_1^+1)$ and ${}^{14} \textrm{C}(2_2^+1)$ have been observed 
by charge exchange reactions \cite{ExpSpec3}. 
They support significant $L_{NN}=2$ component in ${}^{14} \textrm{N}(1_1^+0)$ consistently with the present assignment 
though quantitative reproduction of the $B(\textrm{GT})$ values is not satisfactory in the present calculation. 

For the GT transition between the ground states of ${}^{14} \textrm{C}$ and ${}^{14} \textrm{N}$,
the experimental $B(\textrm{GT}; {}^{14} \textrm{C}(0_1^+1)\rightarrow{}^{14} \textrm{N}(1_1^+0))$ is 
anomalously small as known as a long life problem of ${}^{14} \textrm{C}$.
The suppression of the GT transition of 
${}^{14} \textrm{C}(0_1^+1)\rightarrow{}^{14} \textrm{N}(1_1^+0)$
is partially understood by the $NN$ pair picture in $LS$-coupling scheme that 
${}^{14} \textrm{N}(1_1^+0)$ is not the spin-isospin partner of the ${}^{14} \textrm{C}(0_1^+1)$ but that of ${}^{14} \textrm{C}(2_1^+1)$
because of the large spin-orbit splitting for the $S=1$ $pn$ hole pairs in $L_{NN}=2$. 
It is different from the $A=6$ and $A=10$ systems, in which the lowest $1^+0$ state
is the spin-isospin partner of the ground state of the $N=Z+2$ nucleus.
The GT transition from the $[L_{NN}=0, S_{NN}=0]_{J=0}$ component in ${}^{14} \textrm{C}(0_1^+1)$ to the $[L_{NN}=2, S_{NN}=1]_{J=1}$ component in ${}^{14} \textrm{N}(1_1^+0)$ is forbidden because of the difference $\Delta L_{NN}=2$ in spatial configurations. 
In other words, the GT transition is suppressed because of the $LS$-coupling pair correlation.
Indeed, the calculated $B(\textrm{GT})=0.30$ is factor one smaller than
the sum rule value and less than the half of the $jj$-coupling limit $B(\textrm{GT})=2/3$ for the pure $p_{1/2}^{-2}$ configuration without the pair correlation.
Our result for $B(\textrm{GT}; {}^{14} \textrm{C}(0_1^+1)\rightarrow{}^{14} \textrm{N}(1_1^+0))$ is the same order as those of a NCSM calculation \cite{NCSM_14C} and AMD+VAP calculation \cite{ES_14C} but still largely overestimates the experimental data.
In the present calculation, the $NN$ pairs in ${}^{14} \textrm{C}(0_1^+1)$ and 
 ${}^{14} \textrm{N}(1_1^+0)$ dominantly have $[L_{NN}=0, S_{NN}=0]_{J=0}$
and $[L_{NN}=2, S_{NN}=1]_{J=1}$ components,
respectively, but they are not necessarily ideal $LS$-coupling pairs.
Moreover, $[L_{NN}=2, S_{NN}=1]_{J=1}$ and $[L_{NN}=0, S_{NN}=1]_{J=1}$ are somewhat mixed with each other 
in the obtained ${}^{14} \textrm{N}(1_1^+0)$ and ${}^{14} \textrm{N}(1_2^+0)$. 
As a result of significant mixing of configurations, 
the calculated GT transition ${}^{14} \textrm{C}(0_1^+1)\rightarrow{}^{14} \textrm{N}(1_1^+0)$ does not vanish.
Additional scenarios are required to solve the long-life problem of ${}^{14} \textrm{C}(0_1^+1)$.

In the present analysis, we can understand low-energy spectra of $A=6$, $A=10$, and $A=14$ nuclei
from the $LS$-coupling $NN$ pair picture and assign spin-isospin partners not only for 
the $0^+1$ initial states but also the $2^+1$ initial states
as shown in Fig.~\ref{figure2}.
The spin-orbit splitting of the $J^\pi T=1^+0, 2^+0, 3^+0$ states with $L_{NN}=2$ 
coupled with the intrinsic spin $S=1$ of the $NN$ pair is essential in the spectra of $N=Z=$ odd nuclei. 
In the systematics of the spin-orbit splitting shown in Fig.~\ref{figure2}, 
we can see that the splitting becomes large as $A$ increases. 
It implies that the $LS$-coupling $NN$ pairs feel the stronger spin-orbit mean potential
in heavier systems.

\begin{table*}
\caption{
Expectation values ($\Braket{\bm{S}^2}$ and $\Braket{\bm{L}^2}$) 
of the squared intrinsic spin and orbital angular momentum for
${}^{6} \textrm{Li}$, ${}^{10} \textrm{B}$, and 
${}^{14} \textrm{N}$ obtained by $T\beta\gamma$-AMD+GCM and
${}^{6} \textrm{He}$, ${}^{10} \textrm{Be}$, and 
${}^{14} \textrm{C}$ obtained by $\beta\gamma$-AMD+GCM.}
\label{LLSS}
\begin{ruledtabular}
\begin{tabular}{ccccccccccc}
\multicolumn{4}{c}{$N=Z+2$} &
\multicolumn{7}{c}{$N=Z=$ odd}\\
nuclide&$J_n^\pi T$&$\Braket{\bm{S}^2}$&$\Braket{\bm{L}^2}$&nuclide&$J_n^\pi T$&$\Braket{\bm{S}^2}$&$\Braket{\bm{L}^2}$&$J_n^\pi T$&$\Braket{\bm{S}^2}$&$\Braket{\bm{L}^2}$\\
\hline
${}^{6} \textrm{He}$&$0_1^+1$&0.12 &0.12 &${}^{6} \textrm{Li}$&$0_1^+1$&0.12 &0.12 &$1_1^+0$&1.97 &0.06 \\
 &$2_1^+1$&0.19 &5.65 & &$2_1^+1$&0.20 &5.64 &$1_2^+0$&1.90 &5.75 \\
&&&&&&&&$2_1^+0$&2.00 &5.99 \\
&&&&&&&&$3_1^+0$&2.01 &6.01 \\
\hline
${}^{10} \textrm{Be}$&$0_1^+1$&0.34 &0.34 &${}^{10} \textrm{B}$&$0_1^+1$&0.28 &0.28 &$1_1^+0$&1.94 &0.35 \\
 &$2_1^+1$&0.30 &6.00 & &$2_1^+1$&0.27 &6.04 &$1_2^+0$&1.92 &5.43 \\
&&&&&&&&$2_1^+0$&2.02 &6.49 \\
&&&&&&&&$3_2^+0$&1.97 &7.53 \\
 &$2_2^+1$&0.12 &6.11 & &$2_2^+1$&0.10 &6.08 &$1_3^+0$&1.99 &5.94 \\
&&&&&&&&$2_2^+0$&2.02 &6.61 \\
&&&&&&&&$3_1^+0$&2.05 &7.15 \\
\hline
${}^{14} \textrm{C}$&$0_1^+1$&0.55 &0.55 &${}^{14} \textrm{N}$&$0_1^+1$&0.61 &0.61 &$1_2^+0$&1.94 &0.44 \\
 &$2_1^+1$&0.19 &5.79 & &$2_1^+1$&0.21 &5.83 &$1_1^+0$&1.89 &5.56 \\
&&&&&&&&$2_1^+0$&2.01 &6.07 \\
&&&&&&&&$3_1^+0$&2.02 &6.22 \\
\end{tabular}
\end{ruledtabular}
\end{table*}

\section{Summary and outlook}
\label{summary}
We have studied the Gamow-Teller transitions from $N=Z+2$ neighbors to $N=Z=$ odd nuclei
 in the $p$-shell region
by using $T\beta\gamma$-AMD+GCM.
We have obtained that the strong GT transitions exhausting 
more than 50\% of the sum rule for 
${}^{6} \textrm{He}(0_1^+1)\to{}^{6} \textrm{Li}(1_1^+0)$, 
${}^{10} \textrm{Be}(0_1^+1)\to{}^{10} \textrm{B}(1_1^+0)$, and ${}^{14} \textrm{C}(0_1^+1)\to{}^{14} \textrm{N}(1_2^+0)$.
We have also found the concentration of the GT strengths of the transitions from 
$2_1^+1$ states,
${}^{6} \textrm{He}(2_1^+1)\to{}^{6} \textrm{Li}(1_2^+0,2_1^+0,3_1^+0)$, 
${}^{10} \textrm{Be}(2_1^+1)\to{}^{10} \textrm{B}(1_2^+0,2_1^+0,2_2^+0,3_2^+0)$,
${}^{10} \textrm{Be}(2_2^+1)\to{}^{10} \textrm{B}(1_3^+0,2_1^+0,2_2^+0,3_1^+0)$, and 
${}^{14} \textrm{C}(2_1^+1)\to{}^{14} \textrm{N}(1_1^+0,2_1^+0,3_1^+0)$.
These states connected with the strong GT transitions can be interpreted as ``spin-isospin partner'' states.

For further analysis, we have introduced two-nucleon-pair densities to visualize
$NN$ pair distributions, and found that $S=0,T=1$ $nn$ pairs and 
$S=1,T=0$ $pn$ pairs are dominantly formed in the $N=Z+2$ and $N=Z=$ odd nuclei, 
respectively.
We have studied the 
spin and orbital configurations of the $NN$ pairs in $LS$-coupling scheme
and discussed the behaviors of the $LS$-coupling $NN$ pairs 
in relation to the GT transitions.
The ground states of $N=Z+2$ nuclei, ${}^{6} \textrm{He}(0_1^+1)$, ${}^{10} \textrm{Be}(0_1^+1)$,
and ${}^{14} \textrm{C}(0_1^+1)$, and their partner states, 
${}^{6} \textrm{Li}(1_1^+0)$, ${}^{10} \textrm{B}(1_1^+0)$, and ${}^{14} \textrm{N}(1_2^+0)$, 
have major $L=0$ components, in which both the $NN$ pairs and the core nuclei are in $L=0$ states.
The excited states, ${}^{6} \textrm{He}(2_1^+1)$, ${}^{10} \textrm{Be}(2_2^+1)$, 
and ${}^{14} \textrm{C}(2_1^+1)$, and their partner states have 
dominantly $L=2$ components mainly contributed by the $NN$ rotation around the core, 
whereas ${}^{10} \textrm{Be}(2_1^+1)$ and its spin-isospin partners have 
$L=2$ components with the deformed $2\alpha$ core rotating in $L=2$.
Based on the $LS$-coupling $NN$ pairs, 
the strong GT transitions between spin-isospin partners can be understood 
as spin-isospin-flip phenomena from the $S=0,T=1$ $nn$ pairs in the $N=Z+2$ initial states
to $S=1,T=0$ $pn$ pairs in the $N=Z=$ odd final states.
Namely,
the transitions ${}^{6} \textrm{He}(0_1^+1)\rightarrow{}^{6} \textrm{Li}(1_1^+0)$, ${}^{10} \textrm{Be}(0_1^+1)\rightarrow{}^{10} \textrm{B}(1_1^+0)$ and ${}^{14} \textrm{C}(0_1^+1)\rightarrow{}^{14} \textrm{N}(1_2^+0)$ are spin-flip phenomena of the $NN$ pairs with $[L_\textrm{core}=0, L_{NN}=0]_{L=0}$, whereas 
${}^{6} \textrm{He}(2_1^+1)\rightarrow{}^{6} \textrm{Li}(1_2^+0,2_1^+0,3_1^+0)$, 
${}^{10} \textrm{Be}(2_2^+1)\rightarrow{}^{10} \textrm{B}(1_3^+0,2_1^+0,2_2^+0,3_1^+0)$, and
${}^{14} \textrm{C}(2_1^+1)\rightarrow{}^{14} \textrm{N}(1_1^+0,2_1^+0,3_1^+0)$ are those with $[L_\textrm{core}=0, L_{NN}=2]_{L=2}$.
In the latter cases, the spectra of three final states with $J^\pi = 1^+, 2^+, 3^+$ are split because of the spin-orbit interaction 
for the $S=1,T=0$ $pn$ pairs in $L_{NN}=2$ wave. This spin-orbit splitting plays an important
role in the low-energy spectra of the $N=Z$ odd nuclei.
On the other hand, the spectra of ${}^{10}\textrm{B}(1_2^+0,2_1^+0,2_2^+0,3_2^+0)$ 
partnered with ${}^{10} \textrm{Be}(2_1^+1)$ show small splitting because 
these states have dominant $[L_\textrm{core}=2,L_{NN}=0]_{L=2}$ component, in which 
the spin-orbit interaction does not affect the $S=1,T=0$ $pn$ pairs in $L_{NN}=0$ wave.

In comparison with experimental data, the magnetic moments $\mu$ and the magnetic dipole transition strengths $B(M1)$ are reasonably
reproduced in the present calculation.
Moreover, relatively enhanced $B(\textrm{GT})$ for ${}^{6} \textrm{He}(0_1^+1)\rightarrow{}^{6} \textrm{Li}(1_1^+0)$, ${}^{10} \textrm{Be}(0_1^+1)\rightarrow{}^{10} \textrm{B}(1_1^+0)$, and ${}^{14} \textrm{C}(0_1^+1)\rightarrow{}^{14} \textrm{N}(1_2^+0)$ show consistent features with the present results.
The present calculation also succeeds in describing the concentrations of the GT strengths 
from the $J=2$ excited states: 
${}^{10} \textrm{Be}(2_2^+1)\rightarrow{}^{10} \textrm{B}(3_1^+0)$ and ${}^{14} \textrm{C}(2_1^+1)\rightarrow{}^{14} \textrm{N}(1_1^+0)$.

The present framework, $T\beta\gamma$-AMD+GCM, is a useful tool to systematically study 
the $pn$ pair correlations in $A=6$, $A=10$, and $A=14$ nuclei. 
With this method, we can deal with nuclear deformations 
of the core nuclei and $NN$ pair formation in the same footing. 
This is one of the great advantages superior to 
three-body models with a spherical inert core.

As mentioned above, the strong GT transitions can be understood in terms of the transitions $nn(S=0,T=1)\rightarrow pn(S=1,T=0)$ in $LS$-coupling scheme.
It means that the GT transitions is a good probe to clarify the dynamics of the $pn$ pairs
in $N=Z=$ odd nuclei through the connection with the $nn$ pairs in the neighboring nuclei.
We should comment that anomalous suppression of the 
GT transition ${}^{14} \textrm{C}(0_1^+1)\rightarrow{}^{14} \textrm{N}(1_1^+0)$ is not reproduced in the present calculation and it is still a remaining problem.

In light mass nuclei, the $LS$-coupling $pn$ pairs are formed. However, 
for heavier nuclei, the description of $pn$ pair correlation
in $LS$-coupling scheme is no longer valid 
because $jj$-coupling $pn$ pairs and also the $pn$ pair condensation 
are expected because of the spin-orbit interactions.
Further investigations of $N=Z=$ odd nuclei in a wide mass number region
from light to heavy mass nuclei are required for deeper understanding of $pn$ pair correlations.

\section*{Acknowledgments}
The computational calculations of this work were performed using the supercomputers in the Yukawa Institute for theoretical physics, Kyoto University. 
This work was supported by JSPS KAKENHI Grant Numbers 16J05659 and 26400270.

\end{document}